\documentclass[pre,showpacs,reprint]{revtex4-1}
\usepackage{amsmath,amssymb}
\usepackage{rotating}
\usepackage[utf8]{inputenc}
\usepackage{graphicx} 
\usepackage{epstopdf}

\begin{document}

\title{The ionization of carbon at 10-100 times the diamond density
 and in the 10$^6$ K temperature range.}

\newcommand{\nrc}{National Research Council of Canada, Ottawa, Canada, K1A 0R6}

\normalsize

\author{M. W. C. Dharma-wardana}\email[Email address:\ ]{chandre.dharma-wardana@nrc-cnrc.gc.ca}\affiliation{National Research Council of Canada, Ottawa, Canada, K1A 0R6}

\date{\today}
\begin{abstract}
The behaviour of partially ionized hot compressed matter is critical to the study
 of planetary interiors as well as for nuclear fusion studies.
 A recent quantum 
study of carbon in the 10-70 Gbar range and at a temperature of 100 eV
 used $N$-atom density functional theory (DFT) with
$N\sim 32$-64  and molecular dynamics (MD). This involves band-structure type
electronic  calculations and averaging over many MD generated ion configurations.
The calculated
average number of free electrons per ion, viz., $\bar{Z}$, was 
systematically higher than from a  standard average
 atom (AA)  quantum calculation. To clarify this offset, we examine
(a) the effect of the  self-interaction (SI) error in such estimates
(b) the possibility of carbon being a granular plasma containing Coulomb
crystals. The possibility of  `magic-number' bound states is considered.
The electrical conductivity, pressure, and the compressibility
 of the carbon system are examined. The very low conductivity and the
high $\bar{Z}$ results of DFT-MD point to the existence of carbon in
a complex non-uniform low-conducting dispersed phase, possibly containing
 magic-number Coulomb crystals. The NPA  estimate of $\bar{Z}$, conductivity,
compressibility, and pressure reported here pertain to
the {\it uniform liquid}. 
\end{abstract}
\maketitle

\section{Introduction} 
Astrophysics, planetary science and  high-energy density physics 
rely heavily on theoretical predictions of the equation of state,
transport and radiative properties as the information is not 
usually available experimentally~\cite{GrazDesj14}.
The advent of short-pulse laser
technology has also opened up new regimes of interest, where
theoretical results are needed to interpret new experiments. 
 Computational 
implementations of  $N$-atom density functional theory
(DFT) for electrons moving in the field of $N\sim 100-500$ ions,
 coupled to  molecular dynamics (MD)
simulations for the ions have provided a reliable computational
approach to these types of warm dense matter (WDM)~\cite{VASP,ABINIT}.
This DFT-MD approach will be referred to as the QMD method for brevity.

 QMD provides the properties
of the cluster of $N$-atoms modeled in the simulation cell as a 
periodic solid, but does not provide individual atomic
properties unless additional steps are taken~\cite{Li-Parr86,Bader90,Plage-XRTS15}.
 Such additional steps
usually focus on decomposing density matrices, charge densities,
$N$-atom energies, X-ray
Thomson scattering scattering (XRTS), atomic phase shifts,  $N$ atom
potentials, and such properties into one-atom properties or pair-atom properties, e.g.,
the pair potential~\cite{Stanek21}. However, Bethkenhagen et al.~\cite{BethZbar20}
have recently discussed the decomposition of a transport property,
namely the electrical conductivity of highly compressed hot carbon,
to determine the mean ionization state of the carbon atoms from
a QMD $N$-atom calculation.

 However, the
application of the band-structure based approach to materials like carbon is
  subject to
the well-known band gap error (BGE) which should affect any determination of
the number $\bar{Z}$ of free electrons per ion that has been promoted
over the band gap into the conduction band~\cite{PerdZung81,HybLouie85}.
Thus the fundamental gap of diamond, 5.48 eV, is underestimated by
some 30\% by typical DFT codes, while germanium in its insulator
state ($\bar{Z}=0$) is predicted to be a metal with $\bar{Z}=4$.
 Similar BGE can be expected for the
 1$s$-valance to conduction band energy gap in carbon relevant to
the results of  Ref.~\cite{BethZbar20}. However,  rectifying such
 errors using the  GW method~\cite{HybLouie85}, applied  in the 100
 eV range for densities  $\rho$ at $\sim$ 100 times the normal
density would be computationally very prohibitive. Furthermore,
alternative methods, e.g., average-atom (AA) methods also use
DFT functionals that are subject to a similar self-interaction (SI)
error.

The QMD method uses an electron XC-functional, $F^{xc}_{ee}[n]$ to
reduce the many-electron quantum problem to a theory of non-interacting
electrons, but explicitly deals with $N$-ions without reducing the
$N$-atom problem further. 
Besides the $N$-atom DFT approach of QMD, the `one-atom DFT', i.e.,
'average-atom' (AA) methods directly yield `one-atom' properties
while  taking account of the embedding medium in various
approximations. 

The metallic nature of typical WDMs has been
 exploited to construct  a  theory of these systems
which is computationally very simple and yet flexible and accurate
enough to provide useful results for astrophysically important
low $Z_n$ materials like  C, N, Si etc., as  well as for more general
applications, using a complete density functional
theory of matter where {\it both} the electron subsystem, and the ion
 subsystem are treated by DFT~\cite{DWP1,Pe-Be,ilciacco93,eos95,cdw-pop21}.  
This approach also uses $F^{xc}_{ee}[n]$  to reduce the electron
many-body problem to an effective one-electron problem. In addition,
it reduces the many-ion problem to a one-ion problem using an ion-ion
exchange correlation functional $F^{xc}_{ii}[\rho]$, dependent on the
one-body ion density $\rho(r)$. Here the exchange
component is negligible in typical situations since ions can be
regarded as classical particles. The Hohenberg-Kohn DFT for
classical particles, needed in this approach for electron-ion systems
 is analogous to that for electrons and was
discussed~\cite{Chihara78,EvansDFT79,DWP1} soon after the development of
the DFT treatment of electrons.  The treatment of
effects beyond the Born-Oppenheimer approximation requires an additional 
XC-functional $F^{xc}_{ei}[n,\rho]$.  This is usually neglected but
can be important in special circumstances, and for light  ions~\cite{Furutani90}.
 In this full DFT approach 
the many-electron many-ion problem of many-body interactions is reduced to that
of constructing an object consisting of an ``average single ion of charge $\bar{Z}$
 plus
 $\bar{Z}$ independent electrons'', known as the  neutral pseudoatom (NPA). 
The NPA uses XC-functionals to deal with
three-body and such higher-order inter-particle effects. Thus the NPA is
 a `one-atom DFT' approach which greatly simplifies the $N$-atom
band-structure model  that is implemented in the VASP, ABINIT and similar codes.

 The NPA provides a calculation of the $\bar{Z}$, but it too is subject to
a self-interaction correction (SIC) which is also the cause of the BGE in
the band-structure approach. However, the simplified atomic physics model
 used in the NPA makes it easier to implement a
 SIC based on the Dyson equation~\cite{PDW-Dyson83, PDW-Dyson84}
or the method of Perdew and Zunger~\cite{PerdZung81}. The XC-functionals
for ions used, and the assumption of radial symmetry in the computational
implementation of the  NPA used here restricts the method to uniform
systems.

 Bethkenhagen et al~\cite{BethZbar20} complemented their  QMD results for the
$\bar{Z}$ of carbon with results from the Purgatorio AA
code~\cite{SternZbar07}. Intriguingly, the QMD results
when decomposed to give the average degree of ionization of a single carbon
atom showed a systematic overestimate of $\sim 0.5$ in the value of $\bar{Z}$
compared to the AA $\bar{Z}$.
We have calculated $\bar{Z}$ using a Thomas-Fermi (TF) model, and  our NPA method.
The TF results is almost parallel to the QMD, with a 0.5 offset. The
NPA  finds a somewhat similar but more structured offset.

 The objective of this
 study is to attempt to understand
the magnitude of the likely SIC, and consider other possible reasons for the
differences in the two types of estimates for $\bar{Z}$. Given the higher $\bar{Z}$
found in the QMD, the reported extremely low conductivity from  QMD using a 
Kubo-Greenwood calculation prompts us to consider that the QMD simulations, using
$N=32$ and 64, have converged on a region of the phase space
where the plasma is granular, and possibly made up of Coulomb crystals. We suggest that
various  carbon clusters, each contributing about 0.5 electrons may from
weak $n,l,m$ bound states of a
Coulomb crystal; the principle quantum number $n$ is large enough to envelope
the cluster. Filled shells are energetically favourable and $\bar{Z}$
may increase to achive this.
 These electronic `magic number' states are further stabilized
when clusters with $N$ corresponding to a `structural' magic number (e.g., $N=13$)
 become possible simultaneously. This hypothesis  fits in with the available
 results of the QMD and NPA calculations.

\section{The NPA calculation of $\bar{Z}$}
Detailed discussions of the NPA may be found in several recent 
publications~\cite{Stanek21,cdwSi20}. In this study, the density $\bar{\rho}$,
and temperature $T$ (in energy units) are such that the carbon atom is found to carry 
only the 1$s$ shell of bound electrons, providing a very simple model of an
atom in a plasma. Hence, for simplicity of discussion we develop the two
coupled DFT equations, i.e., for electrons and for the ions that are solved in the NPA,
in the following simplified form where Hartree atomic units ($|e|=\hbar=m_e=1$)
are employed. 

In our `one-atom' DFT model, coupled equations resulting from
the stationary condition of the grand potential $\Omega[n,\rho]$, considered
as a functional of the one body electron and ion densities $n(r),\rho(r)$
are solved. A bare carbon nucleus of charge $Z_n$ is the origin of coordinates
of the uniform system of electrons and ions. Spherical symmetry is applicable,
as we consider a uniform fluid.
The inhomogeneous densities $n(r),\rho(r)$ around the carbon nucleus become the
 `bulk'  densities
$\bar{n},\bar{\rho}$ at large distances $r\to R_c$, where $R_c$ is the
radius of the `correlation sphere'. This is usually
about ten Wigner Seitz radii. Here  $r_{\rm ws}$ is given by
 $r_{\rm ws}=\{3/(4\pi\bar{\rho})\}^{1/3}$. The pair-distribution
functions (PDFs)  $g_{ab}(r), a=e,i$  refer to electrons or ions and 
describe the structure of the environment where the carbon atom is placed.
 Then
\begin{equation}
\rho(r)=\bar{\rho}g_{ii}(r),\;n(r)=\bar{n}g_{ei}(r)
\end{equation}
The PDF $g_{ii}(r)$ will be referred to as $g(r)$ for brevity. 
The grand potential can be written as
\begin{equation}
\Omega=T[N,\rho]+\Omega_e+\Omega_{ei}+\Omega_i .
\end{equation}
Here $T[n,\rho]$ is the kinetic energy functional of a {\it noninteracting}
system having the exact interacting densities. A simplified form for
$\Omega_a$ is given below, assuming a point ion-model $U_{ei}(r)=-\bar{Z}/r$
for the electron-ion interaction. The more complete model, applicable even
 to ion mixtures is found in Ref.~\cite{eos95, Pe-Be} and used in the computations . 
\begin{eqnarray}
\label{omega.eqn}
\Omega_e&=&-\int d{\bf r}\frac{Z_n}{r}n(r)+
\frac{1}{2}\int d{\bf r}d{\bf r'}\frac{n(r)n(r')}{|\bf{r}-\bf{r}'|}\\
        & & \int d{\bf r}F^{xc}_{ee}[n]-\mu_e\int d\bf{r}n(r), \\
\Omega_{ei}&=&-\int d{\bf r}d{\bf r'}\frac{\bar{Z}\rho(r)n(r')}{|\bf{r}-\bf{r}'|}+
\int d{\bf r}F^{xc}_{ei}[n,\rho]
\end{eqnarray}
Note that three-body and higher contributions beyond pair interactions
are all contained in the XC-functionals, and are {\it not} neglected
in the theory.
That is, this one-atom DFT (viz., the NPA) does not use an external
$N$-center potential energy surface due to the ions for the  Kohn-Sham electrons,
as is the case with the $N$-atom DFT deployed in VASP and ABINIT codes.
What we have is the appropriate one-atom mapping of the $N$-atom DFT calculation.

We have used in Eq.~\ref{omega.eqn} a point-ion model $-\bar{Z}/r$ for the electron-ion
interaction of the field ions only for simplicity of
presentation. In actual calculations a local
pseudopotential $U_{ei}(q)=-\bar{Z}V_qM_q$,
where  $V_q=4\pi/q^2$ and a form factor $M_q$ are used. The
form factor, and the corresponding ion-ion pair potential are
also determined self-consistently from the NPA, as discussed
elsewhere~\cite{Pe-Be,eos95,Stanek21}. Non-local forms of the 
pseudopotential have not been found necessary for NPA calculations
for uniform-density  warm dense fluids or for cubic solids.

The ion contribution $\Omega_i$ can be obtained from the above
equations by appropriately replacing $n(r)$ by $\bar{Z}\rho(r)$ if the
ion-electron interaction is modeled by point ions, while also
replacing $F^{xc}$ contributions appropriately. The electron-ion
XC-functional $F^{xc}_{ei}$ is usually neglected in most NPA
calculations, being largely equivalent
to making the Born-Oppenheimer approximation, and neglecting
certain correlation corrections  of the
from $\langle n({\bf r})\rho({\bf r}')\rangle-
\langle n(r)\rangle\langle\rho(r')\rangle$. This is
equivalent to using a `random-phase' approximation for the
electron-ion response function. This is appropriate for
highly compressed {\it uniform} fluids of carbon studied here.
However, such correlations may be important
in dealing with composite carbon grains that form Coulomb
crystals.

The stationary condition on $\Omega$ under functional variation
 $\delta n$ leads
to the usual Kohn-Sham (KS) equation for electrons moving in an effective
potential $U_e(r)$. Functional differentiation with respect to $\delta \rho$
leads to an equation identifiable with the modified hyper-netted-chain
(MHNC) equation if the ion-ion XC-functional is identified with the
hyper-netted-chain (HNC) diagrams and bridge diagrams used for classical
systems. The ions are classical in the regime of study, and there is no
 exchange contribution.
Then the effective classical KS potential for the ions can be
identified with the `potential of mean force', $U_{ii}(r)$ (see appendix to
Ref.~\cite{DWP1}) of classical statistical mechanics.

In actual numerical work, the field ion distribution $\rho(r)=\bar{\rho}g(r)$
 occurring in $\Omega_e(r),\Omega_{ei}(r)$ as well as in the corresponding KS equation
$\delta \Omega/\delta n$ is replaced by $\bar{\rho}g_{cav}(r)$, where $g_{cav}(r)$
is a model ion-ion PDF which is just a spherical cavity of radius $r_{\rm ws}$.
Hence solving the electron KS equation coupled to the ion KS equation
is much simplified, and the only parameter associated with the ion distribution
that has to be varied self-consistently is the ion Wigner-Seitz radius $r_{\rm ws}$
appropriate to a given free electron density $\bar{n}$ given as the input. Thus the
primary input variable is the free electron density, for a given
temperature and nuclear charge.  The equilibrium ion density $\bar{\rho}$ is determined
for each given $\bar{n}$ in this manner, while solving the electron KS
 equation self-consistently, starting from a trial $n(r)$ and $\bar{Z}$.

 The self-consistent solution 
for the continuum and bound state solutions is constrained to satisfy the Friedel
sum rule and verified for satisfying the  $f$-sum rule. The XC-functional used
 for the electron KS equation
is the finite-$T$ XC-functional of Perrot and Dharma-wardana~\cite{PDWXC} within the
local density approximation (LDA). The XC-functional depends on $T/E_F$, where $E_F$
is the Fermi energy of the free electrons. The electron system moves from  a virtually
classical ($T/E_F\sim 2$) electron gas  to a strongly degenerate quantum gas
($T/E_F\sim 0.13$) in the system under study.  A comparison of the finite-$T$
XC-functional used here
 with the parametrization due to Dornheim et al~\cite{Dornheim18} fitted to
 quantum Monte Carlo data showed
 good agreement~\cite{cdw-N-rep19}. The PBE electron XC-functional, i.e., at $T=0$
has been used in the QMD calculations. 

This partial decoupling of the electron Kohn-Sham equation and the ion
 Kohn-Sham equation used in the NPA implementation is
possible because the electron Kohn-Sham equation is found to be only  weakly
dependent on the details of $g_{ii}(r)$ for $r>r_{\rm ws}$. Furthermore, we use
the free electron part of $n(r)$, viz., $n_f(r)=\bar{n}+\Delta n_f(r)$ obtained
 from the KS equation to construct the $\Delta n_f(r)$ that would be obtained
if there were no $g_{cav}(r)$,
using linear response (LR) theory. The corrected $\Delta_f n(r)$ is the response
of a uniform electron fluid (in the presence of a non-responding neutralizing
uniform ion background) to the carbon ion of charge $\bar{Z}$. Hence this corrected
$\Delta n_f(r)$ may be regarded as being  independent of the assumed
form of $g(r)$ in solving the electron KS equation, as long as it satisfied
basic criteria in regard to charge neutrality and the perfect screening
sum rule.     

Once the electron Kohn-Sham equation is solved using $g_{cav}(r)$, we already have
the three quantities $\bar{n},\bar{Z}$ and hence $\bar{\rho}$. We also have the
KS eigenfunctions $\phi_\nu(r)$ and eigenvalues $\epsilon_\nu$, with $\nu=n,l$ for
bound states, and $\nu=k,l$ for continuum states, with $\epsilon_k=k^2/2$, together
with the phase shifts $\delta_{kl}$. These satisfy the Friedel sum rule and the
simple charge neutrality condition $\bar{n}=\bar{Z}\bar{\rho}$. The DFT
calculations used in the NPA also has the usual self-interaction error. Applying
the SIC will make the NPA-estimate of $\bar{Z}$ differ even more from the QMD, unless
the QMD SIC is of the same magnitude. Nevertheless, we examine these corrections before
attempting to discuss other reasons for the offset between our one-atom DFT
results and those of Ref.~\cite{BethZbar20}.

\section{Results for $\bar{Z}$  without self-interaction corrections}

The NPA model described in the previous section does not include self-interaction
corrections. For instance, if the carbon atom has only the $1s$ bound state with
one electron, with an energy $\epsilon_{1s}$, its energy has been calculated with
a KS-potential and a Coulomb repulsion inclusive of its own density. 
 While the effect of this is correctly canceled out in the Hartee-Fock model of 
independent electrons, or in exact-exchange models, the simpler exchange
 correlation functionals do not contain the necessary XC-discontinuity needed
to correct for the self-interaction.
 Hence new functionals that attempt to correct for 
the SIC exist~\cite{Borlido20}, although their success is not as well  established
as for methods based on the Dyson equation~\cite{HybLouie85}.
 In this section we
 first examine the standard results from DFT-MD, the Purgatorio average-atom (AA) model
 as well as the NPA model, where no SIC has been used.
\begin{figure}[t]
\begin{center}
\includegraphics[width=\columnwidth]{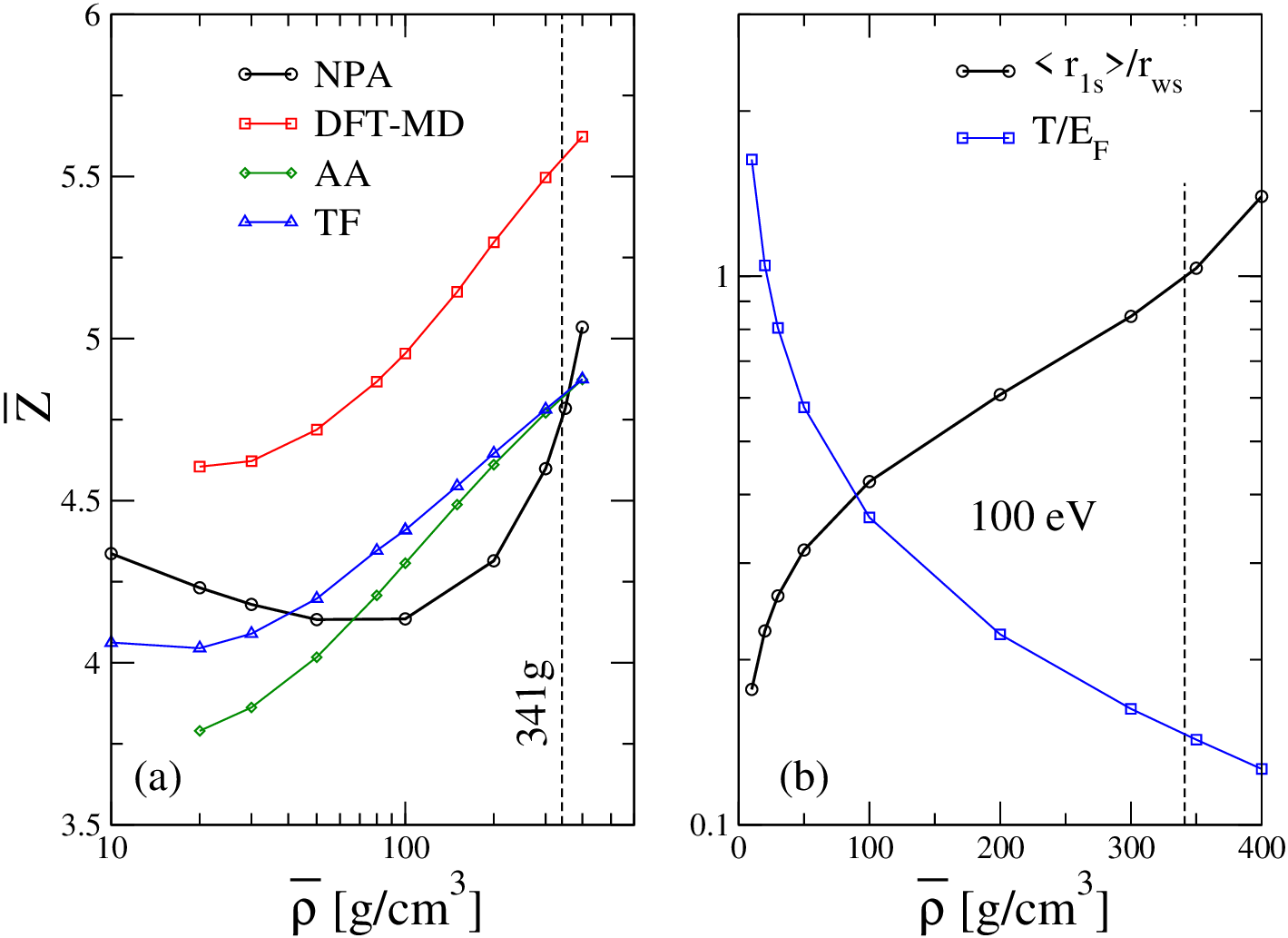}
\caption{
\label{zbar.fig}(color online) Panel (a) $\bar{Z}$ from several models. The
results for the average atom (AA) and for DFT-MD are extracted 
from Ref.~\cite{BethZbar20}. The $\bar{Z}$ from the Thomas-Fermi
model of More et al~\cite{QEOS} are denoted by TF. The dashed vertical
 line labeled 341 g
indicates the density (g/cm$^3$) where the mean radius of the $1s$ orbital
becomes equal to
the ion Wigner-Seitz radius. Panel (b) displays the effective `cooling'
of the system in terms of $T/E_F$, and the 
extension of the $1s$ orbital as the density is increased.
 }
\end{center}
\end{figure}

In fig.~\ref{zbar.fig}, panel (a) we display several  calculations
 of $\bar{Z}$ in the $\bar{\rho}$ range of interest, at 100 eV. No self-interaction
 corrections have  been included. The Thomas-Fermi $\bar{Z}$ used
 in the `Quotidian' equation of state~\cite{QEOS} is also
displayed. The other models are discussed in more detail  below.

\subsection{DFT-MD band-structure model}

In the $N$-atom DFT-MD model (QMD) the $1s$ electrons form a narrow `valance band',
 while the continuum electrons form the `conduction band' for each periodically
repeated solid; i.e., the solid constructed via the MD evolution of the atomic positions
 in the simulation cell. Then an average over an ensemble of band configurations
 generated by MD is carried out, and the average ionization can in
 principle be calculated from the number of electrons per ion in the conduction band.
The $\bar{Z}$ can also be calculated from the one-atom decomposition of DFT-MD
densities given by Plageman et al.~\cite{Plage-XRTS15}. Other methods of increasing
sophistication for decomposing a `molecular property' into `one-atom' contributions 
 are available~\cite{Bader90}.

 A method based on
the $f$-sum rule and the Kubo-Greenwood (KG) conductivity formula 
has been used in Ref.~\cite{BethZbar20}. The method is applicable when there
is a clear separation of the conduction band and the valance band, without
continuum resonances obscuring the bandgap. 
It would be worthwhile examining alternative decompositons of the $N$-atom
properties that depend on $\bar{Z}$ as well, unless a
case can be made for the conductivity being the most satisfactory property for
estimating $\bar{Z}$, even though DFT would require it to be estimated as a 
functional of the density. 
 
The bandgap separating the top of the $1s$ band and the bottom of the
conduction band, as well as the
chemical potential of the electrons enter into the Kubo-Greenwood (KG)
calculation of the conductivity that Bethkenhagen et al.~\cite{BethZbar20} have
used for extracting the number of free electrons per ion, i.e., $\bar{Z}$
in their system. The band-gap error arises partly from the use of the KS
eigenfunctions and eigenvalues whereas the solutions of the Dyson equation
have to be used~\cite{PDW-Dyson83,HybLouie85} in KG. The
 bandgap error in the
regime of densities and at 100 eV is unknown. 

Encouragingly, consistent results have been obtained using the PBE~\cite{PBE96}
functional as well as the SCAN functional~\cite{SCAN13} in the work of
Ref.~\cite{BethZbar20}. Interestingly, the $\bar{Z}$ from QMD is almost
identical to that obtained from the Thomas-Fermi model, except for an
increment $\Delta \bar{Z}$=0.5.

However, it should be noted that the Kubo-Greenwood (KG) formula  when applied
to the determination of the electrical conductivity of liquid-Si  near and above
 its melting point using DFT-MD with the PBE functional or the SCAN
 functional does {\it not} lead to agreement with experiment. 
In the case of Si, there is excellent experimental data for the conductivity
of liquid-Si in a small but technologically important range of densities,
and hence a meaningful comparison is possible~\cite{cdwSi20}. Many physical properties
of liquid-Si  predicted using DFT-MD seem to be
sensitive to the XC-functional used~\cite{Remsing17}.
Unfortunately, there are no experimental conductivities  for
carbon in  this range of densities at 100 eV, or even for
normal densities  near the melting point to test these methods against experiment. 

Hence calculations using other methods, and other codes are useful 
to determine the origin of the spread in the $\bar{Z}$ found between the KG estimate
and  other theoretical approaches for  $\bar{Z}$. 
Here we look at the results from the Purgatorio model and from the NPA model.
  
\subsection{Purgatorio AA model}
In contrast to the NPA, the Purgatorio model confines all the six electrons
of the carbon atom within the WS cell (`ion sphere'), and  hence uses a $\mu$ different
from the non-interacting $\mu^0$. The electron-ion
interaction in the Purgatorio code is set to zero for $r>r_{\rm ws}$. 
 The self-interaction corrections
 should affect AA models in much the same way as for the NPA. According
 to~\cite{SternZbar07}, Purgatorio can be used to yield three estimates of
 ${\bar Z}$ that converge towards the same value at sufficiently high $T$.
 The $\bar{Z}$ given in Fig. 3 of Ref.~\cite{BethZbar20} is stated to be obtained from
 the effective charge at $r=r_{\rm ws}$. The $\bar{Z}$ data, extracted from
 Ref.~\cite{BethZbar20} is displayed in panel (a) of Fig.~\ref{zbar.fig}. It is
seen that the results from the Purgatorio model converge towards the Thomas-Fermi
model as the density is increased.

\subsection{The NPA model}
The NPA model has been used successfully for materials like Al, Si, Li etc., near
their melting points and at  much higher temperatures, unlike many AA models that
seem to be designed mainly with high-$T$ applications in mind.
The NPA boundstates, as well as continuum states extend in the full volume
of the correlation sphere, $R_c\sim 10r_{\rm ws}$.  The NPA differs from many AA models
in that  the electron-ion potential
for $r>r_{\rm ws}$ is not set to zero. 
The electron-ion potential becomes zero only outside the correlation sphere, $r\to R_c$.
The corrections to $\mu^0$  that are taken into
account in elementary non-DFT theories of `continuum lowering' etc., are included in the
KS potentials that become zero only for $r>R_c$, when all the PDFs
$g_{ab}(r), a=e,i$ have attained the asymptotic value of unity.

The carbon NPA has only the single bound state, i.e.,  $1s$,  under the conditions of
the study. Hence the atomic potential is hard and almost point-like.
 Initially, at $\bar{\rho}$ = 20 g/cm$^3$ the electron chemical potential $\mu^0$
is $\sim -18$ a.u; that is, the continuum electrons are classical and hot, with 
$T/E_F\sim 1.7$ at 10 g/cm$^3$. At lower densities, given the negative $\mu$, electrons
ionize easily and $\bar{Z}$ is high. But as $\bar{\rho}$ increases, $\mu$
increases and becomes positive. The ionization is increasingly reduced until about 
100 g/cm$^3$. By then the mean radius of the $1s$ eigenstate of the carbon atom
has reached 42\% of the Wigner-Seitz radius, and reaches 61\% by 200 g/cm$^3$. Then
 the 1$s$ state in the WS sphere begins to leak to the region outside the
ion sphere; strong pressure ionization sets in,
 due to the potential of the ion subsytem modeled by $\bar{\rho}g_{cav}(r)$. 
The figure displays the density of 341g/cm$^3$ when the 1$s$ orbital radius
 becomes equal to $r_{\rm ws}$.

In fact, one may consider that the effect of treating nearest neighbour interactions
more explicitly, not only for this density, but from about
 $\bar{\rho}_{60}$=200g/cm$^3$ may be important. If such nearest-neighbour $1s$-$1s$
interactions are included, we obtain a narrow 1$s$-valance band averaged over
the first peak of the ion-ion PDF. These would appear as transient bonding in
MD simulations. This `valance band'  uses up electrons
in the 1$s$ states, but does not significantly change the estimate of $\bar{Z}$
for the range of $\bar{\rho},T$ studied here. 

\begin{figure}[t]
\begin{center}
\includegraphics[width=\columnwidth]{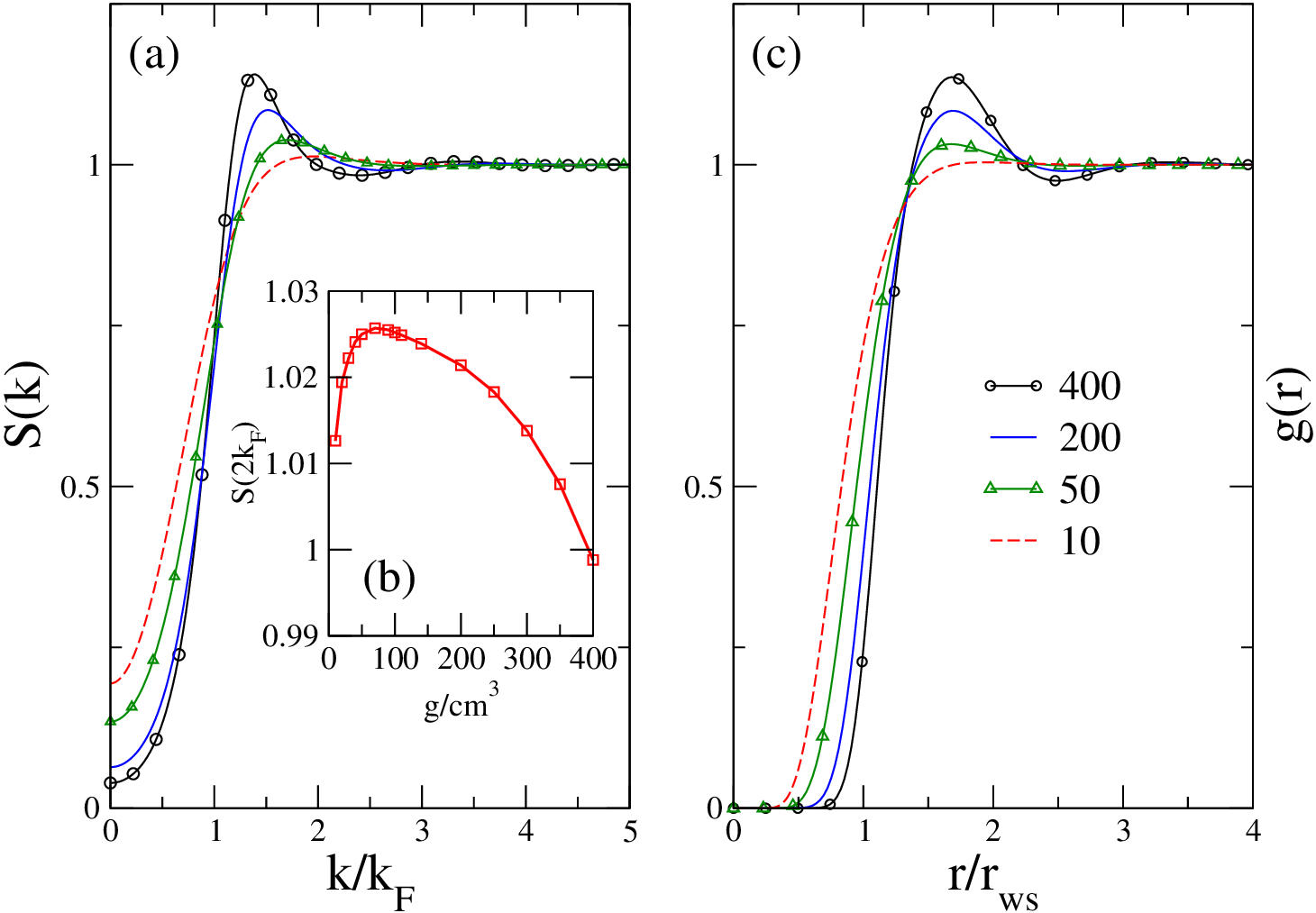}
\caption{
\label{skgr.fig}(color online) Panel (a) The structure factors of
liquid-carbon at 10, 50, 200 and 400 g/cm$^3$ are displayed. Unlike
for normal densities (e.g. diamond $\sim$ 3.5 g/cm$^3$, the first peak
is not split and shows no structure near 2$k_F$. (b) This inset shows the
variation of the value of $S(k)$ at $k=2k_F$ which is determinitive of 
the conductivity for $T/E_F<1$. Panel (c). The C-C pair distribution functions
has no subpeaks or structure due to bonding effects. The position of
the first peak is typical of a densely packed uniform fluid. 
 }
\end{center}
\end{figure}

Nearest-neighbour interactions mediated via the continuum electrons
are treated explicitly in the construction of the pair-potential that is employed to
generate the actual $g(r)$. 
The resulting $g(r)$ differs significantly from $g_{cav}(r)$ outside the
ion sphere. The actual $g(r)$ and the $S(k)$ are shown in Fig.~\ref{skgr.fig}.

The $g(r)$ and $S(k)$ are determined from a hyper-netted-chain calculation (or from MD
if desired) using the ion-ion pair potential defined via the NPA pseudopotential.
 The pseudopotential and the pair-potential used here are given by
\begin{eqnarray}
\label{pseudpot.eqn}
U_{ei}(k)&=&\Delta n_f(k)/\chi(k,\bar{n},T)\\
\label{pairpot.eqn}
V_{ii}(k)&=&\bar{Z}^2V_k+|U_{ei}(k)|^2\chi(k),\; V_k=4\pi/k^2.
\end{eqnarray}
Here $\Delta n_f(k)$ is the Fourier transform of the free-electron density displacement
around the nucleus of the carbon atom placed in the medium with average ion density
$\bar{\rho}$ at $T$. The fully interacting response function of the electron fluid
is denoted by $\chi(k,\bar{n},T)$, and abbreviated to $\chi(k)$, while $V_k$ is the
Coulomb potential. The response function uses a finite-$T$ local field correction
whose $k\to 0$ is chosen to satisfy the compressibility sum rule.

\subsection{The effect of self-Interaction corrections on $\bar{Z}$}
DFT calculations do not normally include a self-interaction
correction. So we include such a correction although we recognize that this
may not resolve the difference between the QMD-prediction
 and the NPA prediction. The inclusion of a SIC makes
 the XC-functional not only
a functional of the density $n(r)$, but also explicitly dependent on the
orbital wavefunction.
\begin{figure}[t]
\begin{center}
\includegraphics[width=\columnwidth]{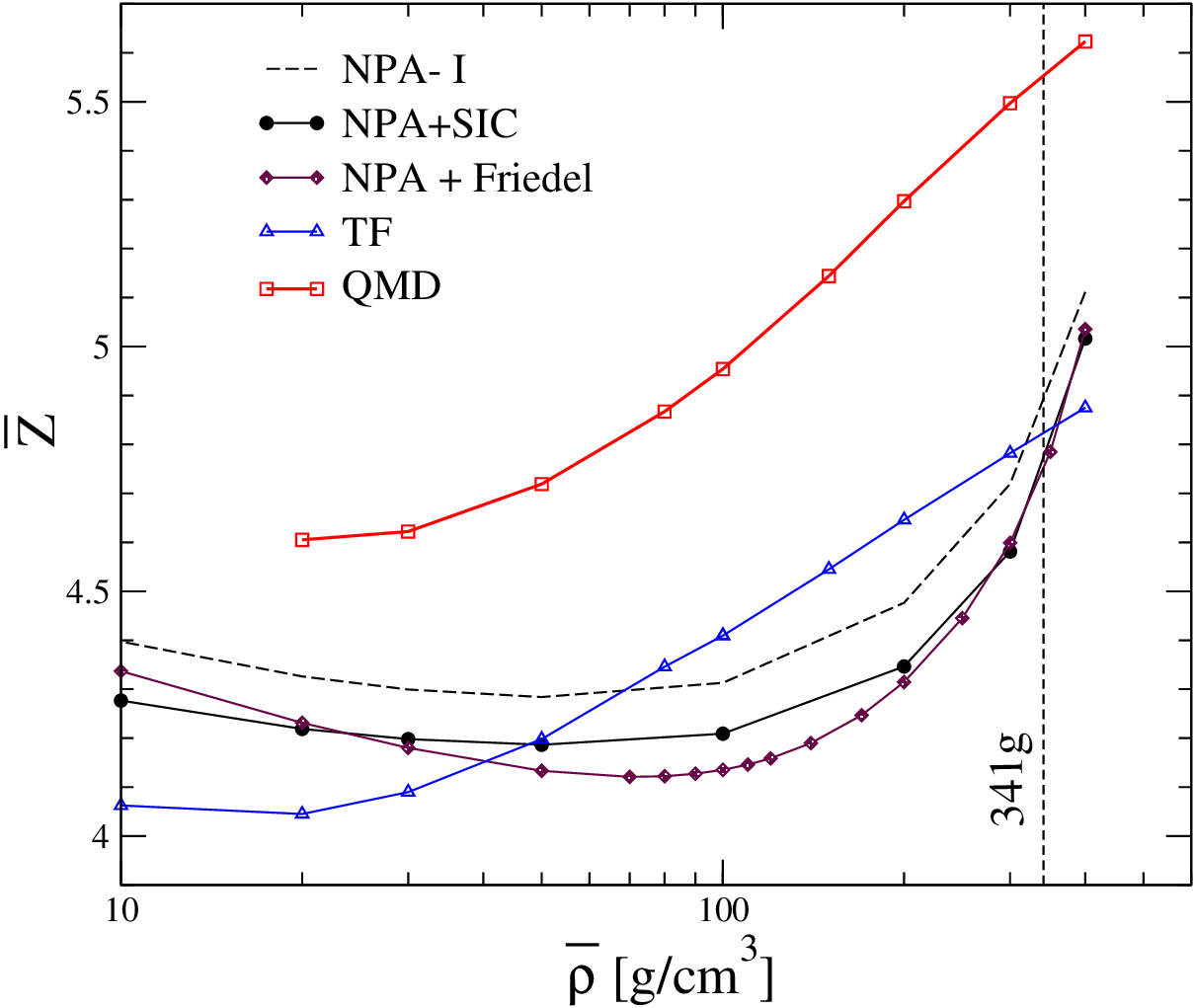}
\caption{\label{zbSIC.fig}(Color online) The NPA estimate of the mean number of
 free electrons per ion, i.e., $\bar{Z}$, is given in successive estimates. NPA-I
is the `naive' estimate, based on the 1$s$ occupation number. This is then
corrected for the  self-interaction error. The calculation imposing
the Friedel sum rule is also given. So the estimate is well-contained and
uncertainties in the NPA-$\bar{Z}$ are definitely smaller than the offset
from the QMD estimate. The Thomas-Fermi $\bar{Z}$ is
shown for comparison. The SC  has been evaluated following
Perdew and Zunger~\cite{PerdZung81}.
 }
\end{center}
\end{figure}

We use the  model of Perdew and Zunger~\cite{PerdZung81} rather than the
Dyson equation approach~\cite{PDW-Dyson83}. In this model, the average
 value of the Coulomb
potential $U(n_\alpha)$ for the electron density in the orbital $\alpha,\sigma$ 
under consideration, and its XC-potential are added together and
subtracted from the uncorrected Kohn-Sham eigenvalue. The expression is
most clearly written in the language of spin-density
functionals.  Given a spin density-functional orbital $\phi_{\alpha,\sigma}$
 containing an electron density
 $n_{\alpha,\sigma}$,  and the fully spin-polarized XC-potential
 $V_{xc}[n_{\alpha\sigma},0]$, the SIC is given as:
\begin{equation}
\label{sic.eqn}
\Delta \epsilon_{sic}=-\langle\phi_{\alpha\sigma}|U(n_{\alpha,\sigma}(r)+
V_{xc}[n_{\alpha\sigma}(r),0]|\phi_{\alpha,\sigma}(r)\rangle
\end{equation}
Here $\phi_{\alpha,\sigma}(r)$ is the eigenfunction obtained by self-consistently
solving the KS equation including the SIC in it. This equation has an orbital
dependency not found in the usual KS equation. So, instead,
 we use the initial (non-SIC) NPA eigenfunction
as an approximate estimate of $\Delta \epsilon_{sic}$. Furthermore, as we use
 a spin-unpolarized
representation, $n_{\alpha,\sigma}=n_\alpha/2$. The zero-$T$ form of the electron XC
functional or its finite-$T$ form can be used with little error since the $T/E_F$
 in the range of study is less than unity for all densities except the lowest.
In Fig.~\ref{zbSIC.fig} we display the SIC-corrected $\bar{Z}$ from the NPA, together
 with the the TF and QMD data (without SIC). The curve marked
 `NPA+Friedel' is the estimate if the Friedel sumrule is strictly imposed. It is
this $\bar{Z}$ that we use in the following calculations. 
Since the SIC
correction is itself an approximation, we may regard the spread among the
 three NPA curves
 as a measure of the uncertainty in our estimate of $\bar{Z}$. 
This
uncertainty is smaller than the offset between the NPA and QMD estimates. 
The effect of the SIC
(band gap error) on the QMD $\bar{Z}$ is likely to be of about the same magnitude as the
SIC in the NPA. Hence we
conclude that the offset is most likely to arise from a deeper underlying 
physical difference.

We do not subscribe to the view of some investigators that $\bar{Z}$ is devoid
of physical meaning, and that it is a convenient fit parameter
 taking various values depending on the physical property investigated, be it the
the conductivity, XRTS, the diffusion coefficient or the opacity etc. 
The $\bar{Z}$ appears in the theory as a valid DFT quantum statistical concept,
 as a Lagrange multiplier for charge neutrality, while $\mu, T$ appear as Lagrange
multipliers for the conservation of particle number and the total energy. The
Kohn-Sham eigenfunctions and energies also have a meaning in that they
are the eigenfunctions and eigenenergies of the non-interacting elections
used in the DFT model. The NPA is also a non-interacting atom constructed from
the interacting ions and electrons of the system.

In all NPA  calculations we use the one unique value of $\bar{Z}$ that satisfies the
atomic physics, the Friedel sum rule, and
the $f$-sum rule, without `fitting' to any physical property.
 It is given by an all-electron atomic physics calculation
of the NPA which incorporates
a sophisticated free energy calculation inclusive of many-body effects. This
is essentially the $\bar{Z}$ that results from a sophisticated
 `Saha' calculation~\cite{eos95}.

The atomic physics of the NPA defines a pseudopotential based on $\bar{Z}$,
and a  pair-potential. They are sufficient to define all the thermodynamic properties
and linear transport properties of the system.
The $\bar{Z}$ is a unique value that applies to all the calculated
properties at a given $\rho,T$.
Optical properties that  involve instantaneous positions (e.g., line
 broadening due to ion microfields) of particles,
 rather than time averaged thermodynamic properties, are outside
its scope unless time-dependent methods are used~\cite{GrimZangw85}. That
 pair-potentials are sufficient for the DFT based NPA approach,
 and that no three-body and higher terms have to be additionally
 included (contrary to the method followed in semi-empirical effective medium theories),
have been discussed and clarified in our recent
 publications~\cite{cdw-N-rep19,cdw-pop21}.

The Perdew-Zunger approach is not easily applied to energy-bands. Even so, one may
expect that if the band-gap error in the DFT-MD calculation were corrected using
the GW method, then a comparable down-shifting of the QMD estimate of $\bar{Z}$
would occur. So, the offset between the two methods remains.
 
In our view, this difference in $\bar{Z}$ arises from the fact that the
 NPA (as well as other well known AA models)
treat a strictly {\it uniform fluid}, while the QMD treats highly
 anisotropic crystals that are thermally  averaged over many
 configurations without any assumption of
radial symmetry. The NPA, together with the Ornstein-Zernike
equation (OZ) and HNC equations can construct an ion-ion XC-functional that can treat
non-uniform structures. The election-ion XC-functional needs to be included
in such calculations. In practice, MD simulations using the NPA pair-potentials
would be more practical in dealing with inhomogeneous systems,
 than using integral equations. In QMD many ionic configurations are
explicitly created and searched.
In our NPA calculations what is obtained is the best possible
{\it uniform liquid} structure even if it be metastable with respect to some lower-energy
solid-state structures (e.g., fullerene structures of carbon, Coulomb crystals)
 that may be possible
 under a given set of $\rho,T$. In contrast, simulations using $N=2n, n=16$
 atoms in the simulation cell may favour icosohedral, folded-graphite forms
(e.g., fullerenes
 and nanotubes in certain ranges of densities)~\cite{MaitiFulleren93}.
 They may include Coulomb-crystals~\cite{BonitzCouCry08, DrewsenCoulobCry15}
that minimize the energy via shell-filling.
 In the following we
present physical properties that support the view that the carbon system revealed
by the QMD study may well be a dispersed complex-solid  phase very different from a
uniform liquid.
   
\section{conductivity}
If the simulations used in Ref.~\cite{BethZbar20} were to generate an average over
many anisotropic crystalline structures, with some components of the conductivity tensor
$\sigma$ having high values, while another principle tensor component is minimally
 conducting, then the averaged conductivity obtained in the KG-calculation will show a very
low conductivity. Alternatively, if the QMD simulation leads to a granular
fluid, then hopping conductivity between grains will be small compared to
that of a uniform metallic fluid.
The fluid-like states studied by the NPA should lead
to a spherically symmetric $\sigma$ with a high conductivity resulting from
 the estimated high $\bar{Z}$. The Ziman formula evaluation~\cite{PDW-Res87}
is implemented within a spherically symmetric $S(k)$ and a radial ($s$-wave)
 pseudopotential. 
   
One may also evaluate the  static conductivity $\sigma$ via the KG equation
using the continuum solutions  of the  NPA model. These too would be for a
 uniform fluid. But they do not contain the scattering from the
 field ions correctly since the NPA (and other
AA-like models) use a spherical cavity, rather than the actual $\rho(r)=\bar{\rho}g(r)$
in modeling the ion distribution. Hence we do not use the KG-estimate of $\sigma$
from the NPA, and instead calculate the conductivity using the
 pseudopotential given in Eq.~\ref{pseudpot.eqn} as well 
as a structure factor generated directly from the NPA calculations.
\begin{figure}[t]
\begin{center}
\includegraphics[width=\columnwidth]{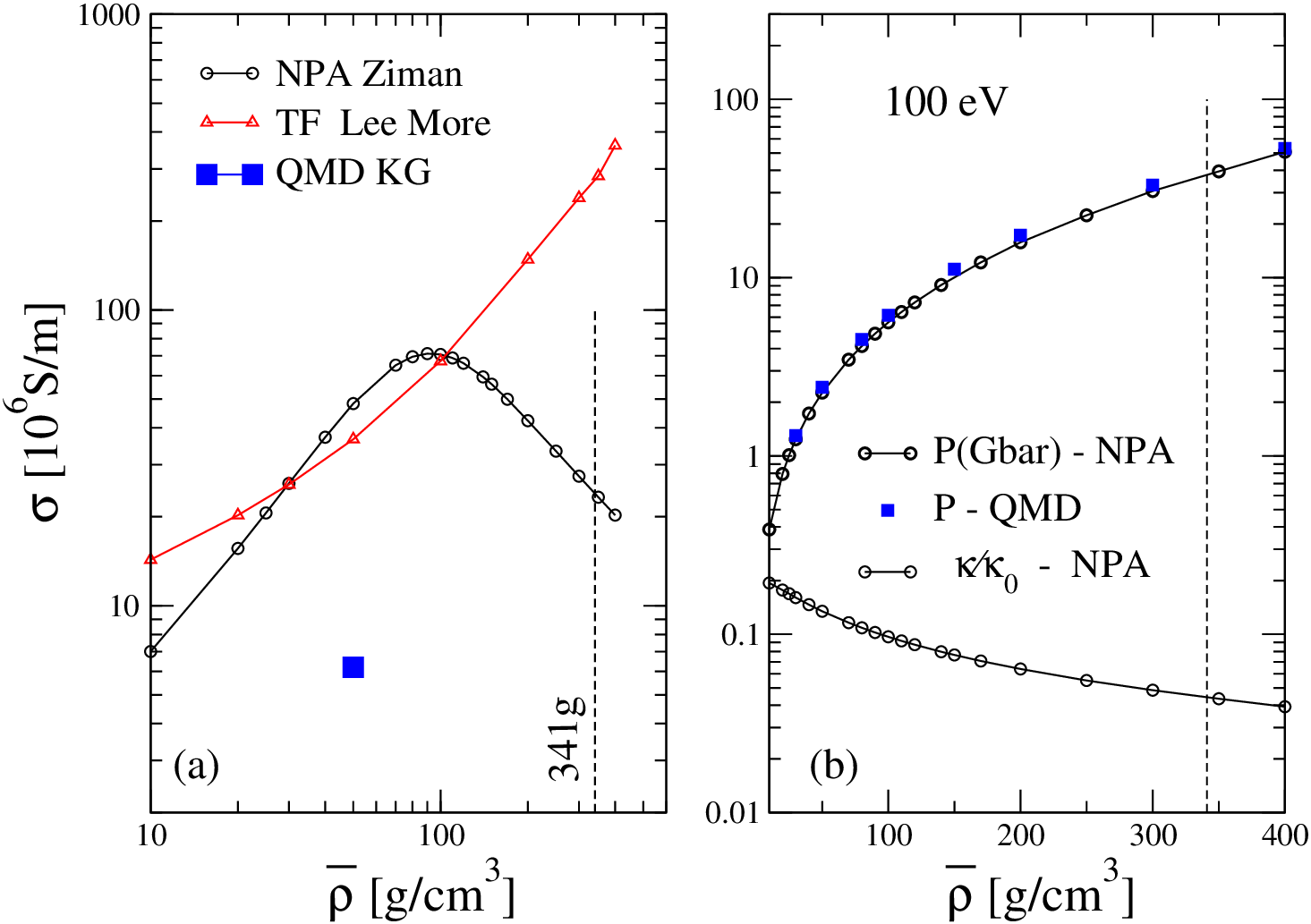}
\caption{\label{cond.fig}(Color online) (a) The electrical conductivity of
carbon evaluated using the NPA+Ziman formula, and using the TF-model
of Lee and More~\cite{LeeMore84}.A single QMD value is given.
 (b) The pressure and the microscopic compressibility $\kappa$
in units of the ideal compressibility $1/T\bar{rho}$.
 This is not derived from
the pressure but taken from the $k\to 0$ limit of $S(k)$. 
The lack of discontinuities
in the pressure or the compressibility rules out any phase transitions
 in the regime
of $T,\bar{rho}$ studied.The QMD pressure, extracted from Fig. 4 of
 Ref.~\cite{BethZbar20} is displayed. Given the higher free-electron
connet of QMD, the NPA pressure being slightly smaller is consistent. 
 }
\end{center}
\end{figure}

The conductivity calculated for the uniform fluid is an order of magnitude higher
than that from the QMD calculation, even though QMD has a higher $\bar{Z}$, and
more free electrons. Unfortunately, only one value of the static conductivity seems
to be reported in Ref.~\cite{BethZbar20}. Interestingly, the
KG estimates of the static conductivity for liquid Li which is also an ion with only a
 small $1s$ core  shows a strong difference between
the Ziman formula estimate and the QMD-KG estimate. In Ref.~\cite{cond3-17} we
 reported that the atomic
 physics obtained from the NPA agreed accurately with the atomic physics of
 the QMD calculation for Li at 0.6 g/cm$^3$ and $T=$4.5 eV,
predicting identical XRTS profiles. But the static conductivities from QMD
 and NPA-Ziman differed by a factor of five~\cite{Witte17}. Lithium  is also
known to form complex structures. The calculation of the conductivity of liquid
sodium near its meting point using the KG-formula is also subject to similar
difficulties unless a very large number of particles is used
 in the simulation~\cite{Pozzo11}.
In the following we consider a particular model of a Coulomb crystal that may
help to explain the reported results from QMD and AA models.

\subsection{Coulomb crystals}
The offset in $\bar{Z}$ between the QMD result, and the results from homogeneous
models (TF,Purgatorio, NPA) is $\sim\Delta \bar{Z}=0.5$. Consider the carbon plasma
at 50 g/cm$^3$. It is
very dense with $r_{\rm ws}=0.863, r_s=0.4018$ a.u., and $\bar{Z}$=4.72, 4.19, 4.13, and 4.02
 in QMD, TF. NPA and Purgatorio respectively. The QMD simulation treats 32 carbon atoms
 as a solid-sate cluster. Since realistic first-principles potentials are being used,
the usual properties of metallic clusters should apply to this simulation as well.
More stable configurations are known as ``magic numbers'', and this may arise from
structural (packing) arrangements, and from electronic ``shell filling'' effects. 
The conditons for these `Coulomb crystals' to form have been discussed by various
authors~\cite{BonitzCouCry08}, although exact conditons can be stated only in the
context of various model systems. Results for hyrogenic systems, e.g., the condition
for the Mott transition on $r_s\sim 1.2$, and similar thresholds can be easily generalized
to systems with $\bar{Z}$. Here we use results from actual calculations and experiments.
The cluster has a nominal radius
\begin{equation}
R_{cl}=r_{\rm ws}N^{1/3}=r_s\{\bar{Z}N\}^{1/3}
\end{equation}
The equations to be solved are very similar to Eq.~\ref{omega.eqn}. The single nucleus
at the center is replaced by a cluster of $N$-ions. In the simplest
`jellium sphere' approximation, this is equivalent to a spherical positive charge
distribution of the form. 
\begin{eqnarray}
\label{clu.eqn}
Q(r)&=&-\bar{Z}\rho(r)\\
 \rho(r)&=&\frac{N}{2R_{cl}}\left[3-\left(\frac{r}{R_{cl}}\right)^2\right],\;
 r\le R_{cl}\\
  &=& -\frac{N}{r},\; r>R_{cl}.
\end{eqnarray}
A more sophisticated form is obtained by introducing a self-consistently determined
 structure factor for the ions
instead of using a jellium sphere. The $\bar{Z}$ included in the above equations
can be optimized is the usual manner, and is subject to the Friedel sum rule and the
$f$sumrule. We do not solve this system as many results of
solutions using DFT, as well as Quantum Monte Carlo results
have been published~\cite{Sotti-DMC01}.
Given that we are dealing with carbon, structures with $N=4$, 13, 55, 58 are
 favoured structurally, with $N$ being the number of ions. Electronic shell filling,
 determined by ``jellium''  magic numbers for the given number of electrons $n_e=\bar{Z}N$
 is determined by the magic numbers  2, 8, 18, 20, 34, 36, 40, 54, 58, 68, 70, 86, 92, 106,
 112, 138, 156,\ldots,
 etc.  Note that 156 electrons  corresponds to about 31-32 carbons atoms with $\bar{Z}=5$,
 or 39 atoms with $\bar{Z}=4$. 

The simulation may produce one, two or several clusters together with individual atoms
 among them  as the MD simulation evolves. The mean density seen by each atom inside
 the cluster
 is quite  oscillatory, and self-compressed by the surface energy of the cluster.
The total energy is also an oscillatory function of $N$, and the structure arises from
 the Kohn-Sham solutions. For such systems, simulations with very large $N$ are
 needed to achieve the properties of the homogeneous system. A small 4-atom carbon cluster,
with a mean ionization $\bar{Z}=4$ has $n_e=16$, and hence achieves the magic number 18 if
it acquires two more electrons and achieves a mean ionization of 4.5, thus registering the
offset of $\Delta Z=0.5$ seen in the QMD cluster calculation versus the uniform-density AA
calculations. Such electrons occupy orbitals weakly localized on the cluster, and such
electrons can contribute weakly to the conductivity by electron hopping from cluster
to cluster. The fact that the simulations is done with periodic boundary conditions does
not change this physical picture.

Many other clusters are possible as there is a choice of magic electron numbers as well
as structural numbers that can combine to give specially stable structures, at
any density in the range explored, with the 32-atom, or 64-atom QMD simulation. A 13-atom
 cluster may have atoms charged $\bar{Z}=4$, having a
 total of 52 electrons, 
and acquire another 6 electrons to achieve the electron magic number 58. Such a structure
is {\it both} structurally and electronically at a magic number, and conforms to HCP, FCC or
icosohedral atomic arrangements. The extra 6 electrons added to complete the `magic shell'
will produce an offset $\Delta\bar{Z}$=6/13=0.46. One may identify a number of other such
likely candidates stabilized by magic-number effects. All these structures may occur
 in a DFT-MD simulation and they would have lower energies than the uniform-fluid
 solution  unless $N$ were very large.

When very large numbers of particles, e.g, $N$=1000 are used  in a
simulation, the effect of such `magic clusters' begins to average out, and bulk-like
behaviour is reached. The physics described by Eq.~\ref{clu.eqn} will approach that of
the uniform `liquid-drop model' as $N\to\infty$ and the oscillatory behaviour will disappear.
 Hence we may conclude that large QMD solutions will show a much smaller offset between
the NPA (AA) predictions, and QMD predictions, both for $\bar{Z}$, and for the
conductivity. On the otherhand, the offset between the QMD results and the average-atom
results may persist, even when the $N$ in the QMD simulations is significantly increased,
 indicating that carbon at 100 eV, at Gbar pressures does not exist as a uniform
 fluid of individual carbon ions.  
\section{Conclusion}
A study of the properties of highly compressed hot carbon using the
`single-atom' DFT approach used in the neutral-pseudo atom model was
undertaken to estimate the extent of self-interaction effects in
this class of plasma models. It was concluded that the offset in the
estimated number of free electrons, $\bar{Z}$ between the average-atom
estimate and the DFT-MD estimate {\it cannot} be explained by self-interaction
errors. The extremely low conductivity predicted by the
 DFT-MD-KG calculations, in comparison to results from the Ziman formula
and from the Lee-More formula strongly suggests that the DFT-MD simulations
using 32 to 64 atoms differed from the average-atom models because the
latter addressed only uniform-fluid solutions. The
possible  existence of `magic-number' stabilized clusters is
hypothesized to be a likely explanation of the observed differences between
the $32-64$-atom DFT-MD simulations, and effective `one-atom' DFT simulations used
in the NPA and Purgatorio codes.    


\end{document}